\documentclass[fleqn,usenatbib]{mnras}

\usepackage{newtxtext,newtxmath}

\usepackage[T1]{fontenc}
\usepackage{ae,aecompl}

\usepackage{graphicx}	
\usepackage{amsmath}	
\usepackage{amssymb}	

\title[Oblate Thin Ocean Waves]{Waves in Thin Oceans on Oblate Neutron Stars} 

\author[B.F.A. van Baal et al.]{
Bart F.A. van Baal,$^{1}$ \thanks{E-mail: bartv.baal@gmail.com}  
Frank R.N. Chambers$^{1}$
and Anna L. Watts$^{1}$
\\
$^{1}$Anton Pannekoek Institute for Astronomy, University of Amsterdam, Postbus 94249, NL-1090 GE Amsterdam, the Netherlands
}

\date{Accepted 2020 June 9. Received 2020 May 29; in original form 2020 February 21}

\pubyear{2020}

\begin{document}
\label{firstpage}
\pagerange{\pageref{firstpage}--\pageref{lastpage}}
\maketitle

\begin{abstract}
Waves in thin fluid layers are important in various stellar and planetary problems. Due to rapid rotation such systems will become oblate, with a latitudinal variation in the gravitational acceleration across the surface of the object.
In the case of accreting neutron stars, rapid rotation could lead to a polar radius smaller than the equatorial radius by a factor $\sim 0.8$. We investigate how the oblateness and a changing gravitational acceleration affect different hydrodynamic modes that exist in such fluid layers through analytic approximations and numerical calculations. 
The wave vectors of $g$-modes and Yanai modes increase for more oblate systems compared to spherical counterparts, although the impact of variations in the changing gravitational acceleration is effectively negligible. 
We find that for increased oblateness, Kelvin modes show less equatorial confinement and little change in their wave vector.
For $r$-modes, we find that for more oblate systems the wave vector decreases. The exact manner of these changes for the $r$-modes depends on the model for the gravitational acceleration across the surface.
\end{abstract}


\newcommand\red[1]{{\color{red}#1}}  

\section{Introduction}

Waves in atmospheres and thin outer layers of stars are very interesting in several astronomical applications. Previous studies have mostly assumed spherical geometry; however, rotating spheroids will become oblate. In particular, accreting neutron stars can rotate up to several hundred times a second which could lead to eccentricities of up to $e = \sqrt{1 - c^2/a^2} \approx 0.5$, or alternatively a flattening $f = 1 - c / a \approx 0.13$, where $a$ is the semi-major axis and $c$ the semi-minor axis of the spheroid. In spite of this, the oblateness of neutron stars is not necessarily taken into consideration for all related phenomena. In this work, we consider the effects of up to second order in the rotation frequency, meaning the neutron star can be modelled as an ellipsoid \citep{chandrasekhar1969ellipsoidal}.

The study of large-scale waves on a sphere comes with a very old and venerable history due to the applications to Earth's oceans and atmosphere. Mathematically, these waves are described by Laplace's tidal equation; a two-dimensional shallow-water model where the ocean is thin relative to the radius of the sphere \citep{pedlosky1987geophysical}. The solutions to these equations consist of different families of waves which have different driving forces. One family are $g$-modes and arise due to the effects of buoyancy and the Coriolis force on displaced fluid packages
\citep[sometimes called gravito-inertial waves, or Poincar\'e waves,][]{gill1982atmospheric,pedlosky1987geophysical,pedlosky2003waves}.
A different kind of low frequency gravity wave which instead arises from the conservation of specific vorticity in combination with the stratification of the ocean and are called Kelvin waves. Such waves are always prograde and as such are sometimes referred to as low frequency prograde waves \citep[][]{unno1989nonradial}.
The $r$-modes, or Rossby waves, arise from the combination of the conservation of specific vorticity combined with the curvature of the surface and the variation of the Coriolis force that comes with such curvatures. These waves can only be found on rotating spheres and are always retrograde. When the radial structure of such modes are considered, they are often given the moniker "buoyant" $r$-modes \citep[][]{heyl2004r,piro2005surface}.
Another wave was discovered due to pioneering work by \cite{yanai1966stratospheric}, called the Yanai mode after this discovery. These modes are sometimes called mixed gravity-Rossby waves and mimic either the $g$-mode at low frequencies or $r$-mode at high frequencies \citep[][]{gill1982atmospheric}.

Previous studies of large scale waves were done assuming a spherical geometry, which is a good approximation for the Earth\footnote{Our own Earth is also oblate due to rotation, although the flattening $f = 3.3\times10^{-3}$ is so low that it is often approximated as a sphere.}. In this work we are going to investigate the effects of oblateness on these families of modes. We use the work of \citet{staniforth2015shallow} who derived a shallow-water model in a non-spherical geometry as was outlined by \citet{white2012consistent}. We solve for linear perturbations in their model and investigate the impact of rotationally induced oblateness on asymptotic approximations and numerical solutions for such perturbations, in a manner similar to \cite{townsend2003asymptotic}.

An astrophysical example where the oblateness might be important is the phenomenon of thermonuclear burst oscillations, first discovered in X-ray observations by \citet{strohmayer1996millisecond}. These oscillations appear due to asymmetric brightness patches on the surface of accreting neutron stars during Type I thermonuclear X-ray bursts and are observed in about $10-20\%$ of all bursts that are observed with high time resolution instrumentation \citep{galloway2008thermonuclear,bilous2019uniform}. Although it is currently unclear what mechanism drives these asymmetries, global wave models were first suggested by \citet{heyl2004r} as a possible explanation. The $m=1$ buoyant $r$-mode is a strong candidate. There have been concerns that this model overpredicts the change in frequency throughout the course the burst \citep{piro2005surface,berkhout2008evolution}, but recent calculations that take into account more up to date nuclear physics and relativistic effects \citep{chambers2019burning,chambers2019relativistic} show a reduced frequency change. A detailed overview of burst oscillations, including their models, potential origins and applications, is given by \citet{watts2012thermonuclear}. 

These recent studies are the main motivation for this work, as the buoyant $r$-mode model for thermonuclear burst oscillations better matches observations when more realistic physics is included. As previously mentioned, accreting neutron stars are expected to be oblate due to their rotation rates which range from $245\,$Hz (IGR~J17511-3057) to $620\,$Hz (4U~1608-522), but effects pertaining to this have not yet been included in the models\footnote{There is one source, IGR~J17480-2446, which has a burst oscillation frequency of $11\,$Hz, but the magnetic field likely plays an important role for this source instead of rapid rotation \citep[see][]{cavecchi2011implications}.}. 

In Section \ref{sec:LTE} we give the mathematical derivation of Laplace's tidal equation for an oblate spheroid, and our model for the gravitational acceleration across the surface; in Section \ref{sec:approximations} we investigate asymptotic approximations for the solutions to these equations and compare to their spherical counterparts. In Section \ref{sec:NS} we apply these new solutions to the case of neutron star burst oscillations by investigating the $r$-modes in detail, before discussing the results in Section \ref{sec:discussion}.

\section{Laplace's Tidal Equations}
\label{sec:LTE}

The strategy for our analysis is based on the work of \cite{townsend2003asymptotic}, that investigated low-frequency pulsation modes in rotating stars. By performing a separation of variables on the perturbations used in the shallow-water equations, it is possible to separate the partial differential equations in spherical-polar coordinates into separate sets of ordinary differential equations for the radial and latitudinal dependence of the perturbation. The important approximation used is the so-called `traditional approximation', which has its origins in geophysics \citep{eckart1960hydrodynamics} but has since been applied to topics ranging from tidal forcing in massive binaries \citep{papaloizou1997non} to investigating gravity modes on rotating neutron stars and their potential link to quasi-periodic oscillations \citep{bildsten1996ocean}. This approximation amounts to neglecting the horizontal component of the angular velocity vector when
 evaluating the Coriolis force \citep[see e.g.][who discuss in which regimes this approximation is valid]{lee1997low}.

The polar (colatitudinal, $\theta$) dependence is governed by Laplace's tidal equations \citep{bildsten1996ocean}, which have a series of eigensolutions first discovered by \cite{hough1898v}. These Hough functions consist of a one-parameter family of solutions in $q=2\Omega/\omega$, where $\Omega$ is the angular velocity of the spheroid and $\omega$ the frequency of the wave, and have an eigenvalue $\lambda$. Only modes with positive $\lambda$ will be considered, although negative values of $\lambda$ are possible as convective modes stabilized by the Coriolis force \citep{lee1986overstable}.

Generally, a numerical approach is necessary to find the solutions to Laplace's tidal equations. However, in the limit that $|q|$ becomes large, asymptotic approximations can be derived. This is well-established in the geophysics literature \citep[see e.g.][]{longuet1968eigenfunctions,gill1982atmospheric}, but the work by \citet{lee1987low} was the first time such approximations were applied to stellar non-radial pulsations, albeit in only the radial direction. It was \citet{bildsten1996ocean} who first presented the asymptotic solutions to the angular equations, and \citet{townsend2003asymptotic} who formally derived this approximation. However, the effects of rapid rotation (large $\Omega$) were not included even though rapid rotation is necessary in order to reach the limit of large $|q|$. 

In this work we consider the effects of oblateness induced by rotation, which alters the problem in two main ways: firstly, the gravitational acceleration will vary with latitude across the surface; the second, and more complex, effect is that by including the oblateness, the radial coordinate $r$ is no longer constant for varying latitude, complicating the separation of variables used to solve the problem. \citet{staniforth2015shallow} derived a set of shallow-water equations in a zonally symmetric geometry, which allows for a latitudinal variation of gravity, known to be significant for even slowly rotating neutron stars \citep{algendy2014universality}.

The separation of the radial and polar coordinates is key in calculating the waves, and by using the parametric ellipsoidal coordinates from \citet{staniforth2015shallow} the equations can be written in such a way that the radial and polar coordinates can be separated once again. This means that a new ODE can be recovered and we can follow many of the steps taken by \citet{townsend2003asymptotic}, although there will be new parameters introduced to track the eccentricity and variation in gravitational acceleration which are present due to the oblateness.

\subsection{Fluid equations including oblateness} 
\label{sec:curvi} 

\newcommand{\Dhor}[1]{\dfrac{ \text{D}_\text{hor} {#1} }{ \text{D} t }}
\newcommand{\azitime}{\mathrm{e}^{i(m\phi + \omega t)}}

\citet{staniforth2015shallow} derive a set of equations appropriate for an incompressible ocean that exists on the surface of a spheroid uniformly rotating about the $z$-axis with angular velocity $\Omega$. These \textit{shallow water equations} describe the fluid flow in terms of azimuthal and latitudinal components of velocity ($u_\phi$ and $u_\theta$ respectively) and the height of the free surface of the fluid layer ($H$, measured from $B(\theta)$ which is a rigid base). They are written:
\begin{subequations}
\begin{equation}
    \Dhor{u_\phi} + \bigg(\dfrac{u_\phi}{a\sin{\theta}} + 2\Omega\bigg) \dfrac{u_\theta\cos{\theta}}{\sigma(\theta)} + \dfrac{1}{a\sin{\theta}} \dfrac{\partial}{\partial\phi}\bigg[ g(\theta) H \bigg] = 0 ,
    \label{eq:basicell_u1}
\end{equation}
\begin{equation}
    \Dhor{u_\theta} - \bigg(\dfrac{u_\phi}{a\sin{\theta}} + 2\Omega\bigg) \dfrac{u_\phi\cos{\theta}}{\sigma(\theta)} + \dfrac{1}{a\sigma(\theta)} \dfrac{\partial}{\partial\theta}\bigg[ g(\theta) H \bigg] = 0 ,
    \label{eq:basicell_u2}
\end{equation}
\begin{equation}
    \Dhor{\left[ H - B(\theta) \right]} + \dfrac{H - B(\theta)}{a\sin{\theta}} \bigg[ \dfrac{\partial u_\phi}{\partial \phi} + \dfrac{1}{\sigma(\theta)} \dfrac{\partial}{\partial \theta}\big( u_\theta\sin{\theta} \big) \bigg] = 0 ,
    \label{eq:basicell_continuity}
\end{equation}\label{eq:nonlin_sw}
\end{subequations}
where 
\begin{equation}
    \Dhor{} \equiv \dfrac{\partial}{\partial t} + \dfrac{u_\phi}{a\sin{\theta}}\dfrac{\partial}{\partial \phi} + \dfrac{u_\theta}{a \sigma(\theta)}\dfrac{\partial}{\partial\theta} 
    \label{eq:basicell_Dhor}
\end{equation}
is the horizontal material derivative\footnote{\citet{staniforth2015shallow}, equations (48)-(51), are written in term of latitude, $\theta_{\text{l}}$, whereas we use to co-latitude, $\theta$. Converting between these coordinates introduces a sign change in the north-south velocity. Using their notation, the latitudinal velocity is related to the co-latitudinal velocity as $u_2 = - u_\theta$.}.
The eccentricity of the ellipse is $e \equiv \sqrt{(1 - c^2 / a^2)}$ where $a$ and $c$ are the semi-major and semi-minor axes of the spheroid respectively. The function $\sigma(\theta)$ depends on co-latitude and the eccentricity as $\sqrt{1 - e^2 \sin^2{\theta}}$. The surface $B(\theta)$ will be assumed to form a geopotential surface, thus $ \partial_{\theta} \left[ g(\theta) B(\theta) \right] = 0$. See \cite{staniforth2015shallow} for the full set of assumptions used in deriving these equations.

The gravitational acceleration on the surface of the ellipsoid, $g(\theta)$, is not constant, which is a departure from the shallow water equations on the surface of a sphere. The dependence of the gravitational acceleration on the co-latitude is discussed further in Section \ref{sec:gravintro}. The product of the gravitational acceleration and the height of the free surface, $g H$, is constant along the surface of the ellipsoid; this result is related to the fact that the pressure is assumed to be constant along the free surface.

We solve for perturbations of the form:
\begin{subequations}
\begin{align}
    g H &= g B + g H_0 + \epsilon P_p(\theta) \azitime , \\
    \sin\theta u_\theta &= i \epsilon \omega P_\theta(\theta) \azitime , \\
    \sin\theta u_\phi &= \epsilon \omega P_\phi(\theta) \azitime ,
\end{align}\label{eq:curvi_gensol}%
\end{subequations}
where $\omega$ is the wave frequency, $m$ denotes the azimuthal wave number (which is constrained to integer values), and $\epsilon$ is a small factor used to keep track of perturbed terms. $H_0$ is the height of the fluid layer at rest. The Lagrangian fluid displacement is related to the fluid velocity perturbation as $\boldsymbol{u} \equiv \text{D}_\text{hor} {\boldsymbol{\xi}}/ \text{D} t$. The pressure along the free surface is related to the height as $p = p_0 + \rho g H$ where $p_0$ is a reference surface. The wave propagates either prograde or retrograde depending on the signs of $m$ and $\omega$. Prograde (retrograde) motion corresponds to $m \omega < 0$ ($m \omega > 0$).

We apply the perturbations in Equations (\ref{eq:curvi_gensol}) to Equations (\ref{eq:nonlin_sw}), retaining terms linear in $\epsilon$.
We define the parameters $q \equiv 2 \Omega / \omega$ and $\lambda \equiv a / (H g \rho)$\footnote{The eigenvalue $\lambda$ is related to the wave vector $k$ as $k\sim\lambda^{1/2} / R$ in spherical geometry.} and use the substitution $\mu \equiv \cos(\theta)$ and $\mathcal{D} \equiv (1 - \mu^2) \text{d} / \text{d} \mu$ to find the following set of ordinary differential equations: 
\begin{subequations}
\begin{equation}
    \mathcal{D}P_p = -\sigma P_\theta - q\mu P_\phi ,
    \label{eq:curvi_houghTheta}
\end{equation}
\begin{equation}
    mP_p = -P_\phi - q\mu \dfrac{1}{\sigma}P_\theta ,
    \label{eq:curvi_houghPhi}
\end{equation}
\begin{equation}
    \lambda \sigma (1-\mu^2) P_p = -\sigma m P_\phi + \mathcal{D}P_\theta - \dfrac{P_\theta}{g}\mathcal{D}g .
    \label{eq:curvi_houghContinuity}
\end{equation}
\end{subequations}
Note that these equations reduce to their spherical counterpart when $e = 0$ and the gravitational acceleration becomes constant \citep[equations (17)-(19) in][]{townsend2003asymptotic}. The quantity $P_\phi$ can be eliminated and the equations become:
\begin{equation}
    \bigg( \mathcal{D} - mq\mu \bigg)P_p = \sigma\bigg(\dfrac{q^2\mu^2}{\sigma^2} -1\bigg)P_\theta,
    \label{eq:curvi_houghP_ODE}
\end{equation}
\begin{equation}
    \bigg( \mathcal{D} + mq\mu \bigg)P_\theta - P_\theta\mathcal{D}\ln{g} = \sigma \bigg[ \lambda(1-\mu^2) - m^2 \bigg] P_p.
    \label{eq:curvi_houghTheta_ODE}
\end{equation}
It is possible to eliminate $P_\theta$ and find one second order ordinary differential equation for $P_p$ which reduces to Laplace's tidal equation in the limit that $e = 0$ and the gravitational acceleration becomes constant. Further setting $q = 0$, the equation reduces to the associated Legendre equation.

\subsection{Variation in gravitational acceleration} \label{sec:gravintro}
Including oblateness introduces an extra term in Equation (\ref{eq:curvi_houghTheta_ODE}) compared to the equation in spherical geometry. We require a model for the variation in gravitational acceleration with latitude across the surface of the spheroid. Both the eccentricity $e$ and any model for $g(\theta)$ are functions of the mass, radius, rotation rate and equation of state of the spheroid. 

Any model for $g(\theta)$ must be symmetrical about the equator. We choose to model this function as a polynomial in $\cos \theta$,
\begin{equation}
	g(\mu) \equiv g_\mathrm{E}(1 + \chi\mu^2),
    \label{eq:grav_eq}
\end{equation}
where $ g_\mathrm{E}$ is the gravitational acceleration at the equator, and $\chi$ is a new parameter which describes how gravity varies with latitude. The gravitational acceleration at the pole is $g_{\text{E}} (1 + \chi)$.
Through this equation, instead of dealing with parameters for radii, masses, rotation rates and equations of state, we can describe our variations through $e$ and $\chi$ alone. This adds only two extra parameters to the shallow-water equations. Equation (\ref{eq:grav_eq}) is just one way of describing $g(\theta)$, and more complicated choices are certainly possible.

We now continue by making asymptotic approximations, similar to \citet{townsend2003asymptotic}, while including these new parameters that account for oblateness.

\section{Results} \label{sec:approximations}

In this section we find asymptotic approximations for the eigenvalues and eigenfunctions of Equations (\ref{eq:curvi_houghP_ODE}) and (\ref{eq:curvi_houghTheta_ODE}) for large spin parameter and a range of values for eccentricity. We compare these approximations to their spherical counterparts and numerical solutions.

\citet{yoshida1960theory} recognised that for large values of the spin parameter, $q$, certain families of modes become equatorially trapped due to the Coriolis force. This equatorial trapping was the principal property used by \cite{townsend2003asymptotic} to derive asymptotic solutions, where it was noted that the spin parameter always appears together with the latitudinal coordinate in Laplace's tidal equation. Since eigenfunctions must remain finite over the whole range of latitude, they can only be significantly different from zero around the equator where terms $q \mu$ are or order unity. Similar arguments are true for these equations when the effect of oblateness is included. This leads to the simplification, $\mathcal{D} P_i \approx \mathrm{d} P_i / \mathrm{d} \mu$. 

\subsection{Asymptotic approximations for $g$-modes, Yanai modes and $r$-modes}

By assuming that $\lambda$ is appreciably different from $m^2$, we may also drop terms of order $\lambda \mu^2 P$ on the right hand side of Equation (\ref{eq:curvi_houghTheta_ODE}). These two simplifications lead to the equations:
\begin{equation}
    \bigg( \dfrac{\text{d}}{\text{d}\mu} - mq\mu \bigg)P_p = \sigma\bigg( \dfrac{q^2\mu^2}{\sigma^2} - 1 \bigg)P_\theta,
    \label{eq:Pp_to_Ptheta_curvi}
\end{equation}
\begin{equation}
    \bigg( \dfrac{\text{d}}{\text{d}\mu} + mq\mu \bigg)P_\theta - P_\theta\dfrac{\text{d}}{\text{d}\mu}\ln{g} = \sigma\bigg( \lambda - m^2 \bigg)P_p.
    \label{eq:Ptheta_to_Pp_curvi}
\end{equation}
$P_p$ can be eliminated to obtain a second-order differential equation for $P_\theta$:
\begin{equation}
\begin{split}
   \dfrac{\text{d}^2 P_\theta}{\text{d}\mu^2}
   -\bigg( \dfrac{\text{d}\ln{g}}{\text{d}\mu} + \dfrac{e^2\mu}{\sigma^2} \bigg)\dfrac{\text{d}P_\theta}{\text{d}\mu}
   + \bigg( mq - \lambda q^2\mu^2 + \sigma^2\lambda - \sigma^2m^2 \\ - \dfrac{e^2\mu^2mq}{\sigma^2} -  \dfrac{\text{d}^2\ln{g}}{\text{d}\mu^2} + mq\mu\dfrac{\text{d}\ln{g}}{\text{d}\mu} + \dfrac{e^2\mu}{\sigma^2}\dfrac{\text{d}\ln{g}}{\text{d}\mu} \bigg) P_\theta = 0.
   \label{eq:Ptheta_ODE_gravityfunc}
\end{split}
\end{equation}
We now implement our model for the gravitational acceleration across the surface of the star, discussed in Section \ref{sec:gravintro}. Again, we take advantage of the fact that the eigenfunction is only appreciably different from zero in the region around the equator. This leads to the approximations $\text{d}_\mu \ln{g} P_\theta \approx 2\chi\mu P_\theta$ and $\text{d}_\mu^2 \ln{g} P_\theta \approx 2 \chi P_\theta$. For the same reason, we also make the simplification that $\sigma(\mu) P_i \approx \sqrt{1 - e^2} P_i$, which can be used to simplify Equation (\ref{eq:Ptheta_ODE_gravityfunc}):
\begin{equation}
\begin{split}
    \dfrac{\text{d}^2P_\theta}{\text{d}\mu^2} - \mu\bigg( 2\chi + \dfrac{e^2}{\sigma^2} \bigg)\dfrac{\text{d}P_\theta}{\text{d}\mu} + \bigg( mq - \lambda q^2\mu^2 + \sigma^2\lambda - \sigma^2m^2 \\ - \dfrac{e^2\mu^2mq}{\sigma^2} + 2\chi mq\mu^2 + 2\chi\dfrac{e^2\mu^2}{\sigma^2} -2\chi \bigg)P_\theta = 0.
    \label{eq:Ptheta_ODE_gravapproximated}
\end{split}
\end{equation}
In order to simplify the analysis of this equation, we introduce the following variables:
\begin{equation}
    L^2 \equiv \lambda,
    \label{eq:lamL_dub}
\end{equation}
\begin{equation}
    \Upsilon \equiv \dfrac{e^2}{\sigma^2}mq - 2\chi mq - 2\chi\dfrac{e^2}{\sigma^2},
    \label{eq:UpsL}
\end{equation}
\begin{equation}
    \alpha \equiv (L^2 q^2 + \Upsilon)^{1/4}\mu,
    \label{eq:alphaL}
\end{equation}
\begin{equation}
    A \equiv \dfrac{mq + \sigma^2(L^2   - m^2) - 2\chi}{(L^2 q^2 + \Upsilon)^{1/2}},
    \label{eq:AL}
\end{equation}
\begin{equation}
    E \equiv - \dfrac{2\chi + \frac{e^2}{\sigma^2}}{(L^2 q^2 + \Upsilon)^{1/2}}.
    \label{eq:EL}
\end{equation}
With these new definitions comes an extra constraint; $\alpha$ must be real and thus $L^2q^2 \geqslant -\Upsilon$. With the correct substitutions and replacing $\mu$ as a variable with $\alpha$, the ODE can be written as: 
\begin{equation}
    \dfrac{\text{d}^2P_\theta}{\text{d}\alpha^2} + \alpha E \dfrac{\text{d}P_\theta}{\text{d}\alpha} + (A - \alpha^2)P_\theta = 0.
    \label{eq:Ptheta_alpha_ODE}
\end{equation}
In the case that $E=0$ this equation reduces to the time-independent Schr\"odinger equation for a quantum harmonic oscillator for a particle trapped in a potential well \citep[][]{arfken1999mathematical}. In a similar manner, waves are trapped in the region around the equator; $E$ and $\Upsilon$ are responsible for the effect of oblateness which alters this potential.

A general solution to Equation (\ref{eq:Ptheta_alpha_ODE}) is given by a combination of a Hermite Polynomial ($H_s$) and Kummer's Confluent Hypergeometric Function ($_1F_1$). Although there is a relation between $H_s$ and $_1F_1$ , fortunately we need not worry about this because 
the boundary conditions of the problem simplify this solution dramatically; either $P_p$ or $\mathrm{d} P_p / \mathrm{d} \mu$ are zero at the equator depending on the parity of the mode \citep[][]{bildsten1996ocean}. Since $_1F_1$ will not be zero at these points, this part of the general solution must vanish in order to satisfy the boundary conditions.
Thus, the general solution is given by,
\begin{equation}
    P_\theta(\alpha) = \text{e}^{-\frac{1}{4}\big(\sqrt{E^2+4}+E\big)\alpha^2} H_s \bigg(\dfrac{\alpha}{\sqrt{2}}\sqrt[4]{E^2+4}\bigg).
    \label{eq:Ptheta_alpha_ODE_nokummer}
\end{equation}
With this specific solution comes the integer index $s$ which is given by: 
\begin{equation}
s = -\frac{\sqrt{E^2+4}+E-2A}{2\sqrt{E^2+4}},
\label{eq:hermite_constraint}
\end{equation}
where the integer $s\geqslant 0$. From Equation 
(\ref{eq:hermite_constraint}) the following relation can be recovered:
\begin{equation}
    A^2 - AE - s(s+1)E^2 = (2s+1)^2.
    \label{eq:curvi_s_AE_rel}
\end{equation}
Combining this relation with Equations (\ref{eq:AL}) and (\ref{eq:EL}), an expression quadratic in $\lambda$ can be obtained. From there it is possible to find the two roots for $\lambda_\pm$:
\begin{equation}
\begin{split}
    L^2_\pm &= -\dfrac{mq\sigma^2 - m^2\sigma^4 - 2\chi\sigma^2}{\sigma^4} - \dfrac{2\chi\sigma^2 + e^2}{2\sigma^4} \\ &+ \dfrac{(2s+1)^2q^2}{2\sigma^4}\Bigg\{ 1 \pm \bigg[ 1 - \dfrac{4(mq\sigma^2 - m^2\sigma^4 - 2\chi\sigma^2)}{(2s+1)^2q^2} \\ & - \dfrac{2(2\chi\sigma^2 + e^2)}{(2s+1)^2q^2} + \, \dfrac{4\sigma^4\Upsilon}{(2s+1)^2q^4} + \dfrac{(2\chi\sigma^2 + e^2)^2}{(2s+1)^2q^4} \bigg]^{1/2} \Bigg\},
\label{eq:Lsquared_curvi}
\end{split}
\end{equation}
and by using a Taylor expansion these roots can be approximated as
\begin{equation}
    \lambda_+ = \dfrac{(2s+1)^2q^2}{(1-e^2)^2} - 2\bigg[\dfrac{mq}{(1-e^2)} - m^2 - \dfrac{2\chi}{(1-e^2)}\bigg],
    \label{eq:lambda_gmodes}
\end{equation}
and
\begin{equation}
    \lambda_- = \dfrac{\big[mq - m^2(1-e^2)\big]^2 }{(2s+1)^2q^2} + \dfrac{mq\bigg[\dfrac{e^2}{(1-e^2)} - 2\chi\bigg]\big[1+(2s+1)^2\big]}{(2s+1)^2q^2}.
    \label{eq:lambda_rmodes}
\end{equation}
Equations (\ref{eq:lambda_gmodes}) and (\ref{eq:lambda_rmodes}) reduce to the forms given by \cite{townsend2003asymptotic} in spherical geometry.

The two different branches of eigenvalues are associated with different types of waves. The $\lambda_+$ branch are known as $g$-modes, sometimes called Poincar\'e waves \citep[][]{gill1982atmospheric}, while the $\lambda_-$ branch is associated with $r$-modes, since all valid solutions must be non-axisymmetric and retrograde \citep[][]{saio1982r}. \citet{townsend2003asymptotic} furthermore finds solutions for a special case where $s=0$, sometimes called the mixed gravity-Rossby wave since it can create both types of waves depending on whether the mode is retrograde or prograde. However, these kinds of waves were discovered by \citet{yanai1966stratospheric} and as such are also known as Yanai waves in honour of this discovery. In the limit of large $|q|$, the Yanai modes are best approximated as $g$-modes with $s=0$ in the case of either prograde or retrograde propagation.

\subsection{Gravitational acceleration connected to the eccentricity} 
\label{ssec:eigenvalues}

We now connect the value of the parameter dictating the gravitational acceleration across the surface of the star, $\chi$, with the eccentricity, $e$. We make specific assumptions for the case of neutron star oceans using the models established by \citet{algendy2014universality}.
In order to investigate a relation between $\chi$ and $e$, we explore a range of values for compactness, angular velocity, mass and radius\footnote{See Equations (20) and (49) in \citet{algendy2014universality} for radius and gravitational acceleration, respectively.}.

We are particularly interested in neutron stars with thermonuclear burst oscillations, which are known to have spin frequencies as high as $620\,$Hz. For neutron stars rotating at these high frequencies, $e$ and $\chi$ follow the approximate relation:
\begin{equation}
	\chi = \dfrac{4}{3}e^2.
    \label{eq:realistic_chi_e_rel}
\end{equation}
Assuming a neutron star with equatorial radius between $10$ and $14.5\,$km, and mass between $1.3$ and $2.1\,\textrm{M}_\odot$ for a variety of equations of state, we find $e$ lies in the range $e=0.08 - 0.23$ or $e=0.22-0.57$ for a rotation rate of $245\,$Hz and $620\,$Hz, respectively. In order to more thoroughly investigate the effect of a larger variation of gravitational acceleration from the equator to the pole, we will also consider the relation $\chi=2e^2$.

\begin{figure}
	\includegraphics[width=\linewidth]{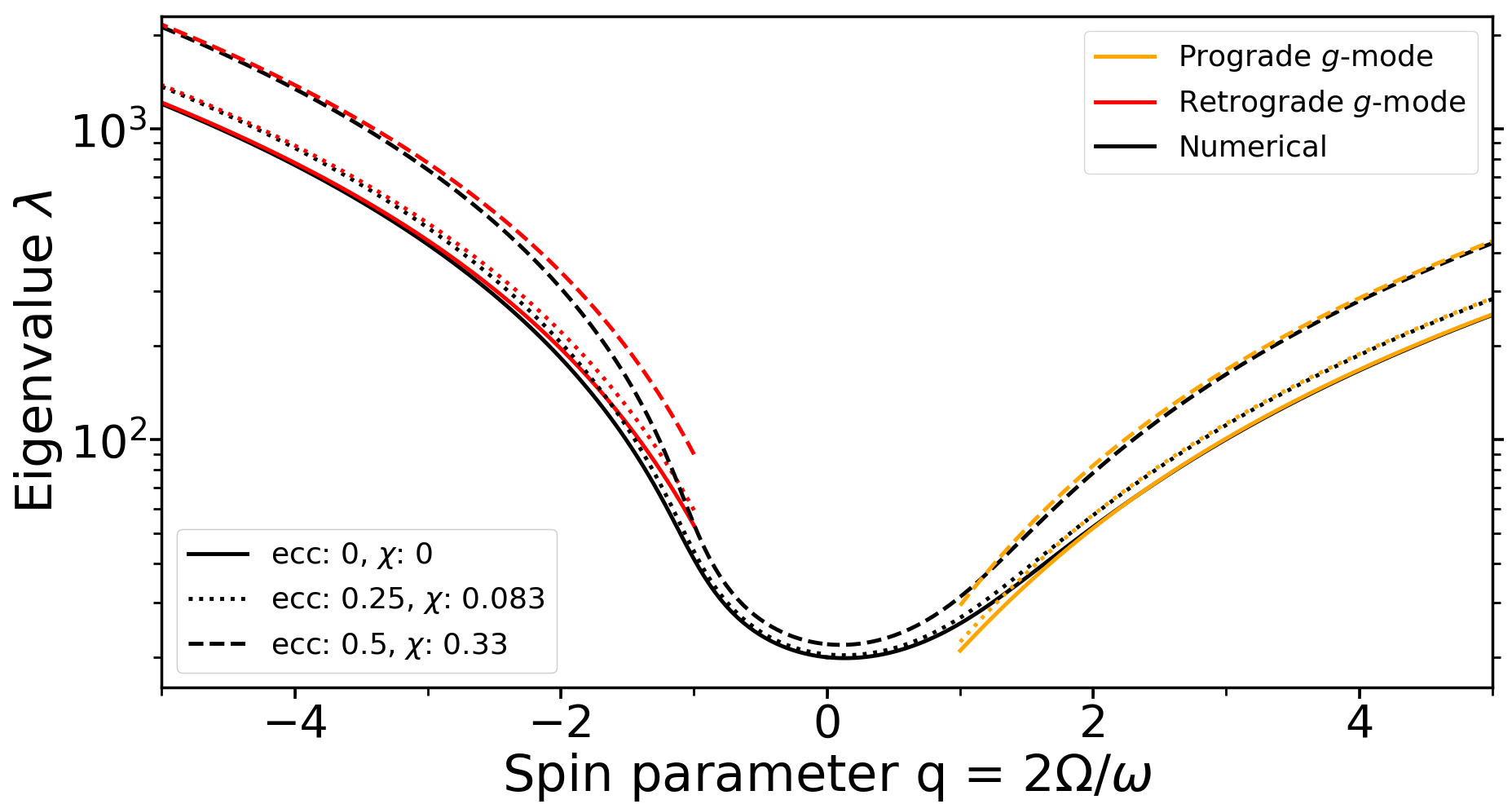}
    \caption{A comparison between the numerically calculated (black) and approximate  (in colour) solutions to the $m=-2$ prograde and retrograde waves with $k=2$ for different eccentricities. Orange corresponds to a prograde $g$-mode and red to a retrograde $g$-mode. The solid lines correspond to $e=0$, the dotted lines to $e=0.25$ and the dashed lines to $e=0.5$, with the relation between $e$ and $\chi$ as given by Equation (\ref{eq:realistic_chi_e_rel}). The analytic approximations are given by Equation (\ref{eq:lambda_gmodes}) for both modes.}
    \label{fig:k2proretro}
\end{figure}

\begin{figure}
	\includegraphics[width=\linewidth]{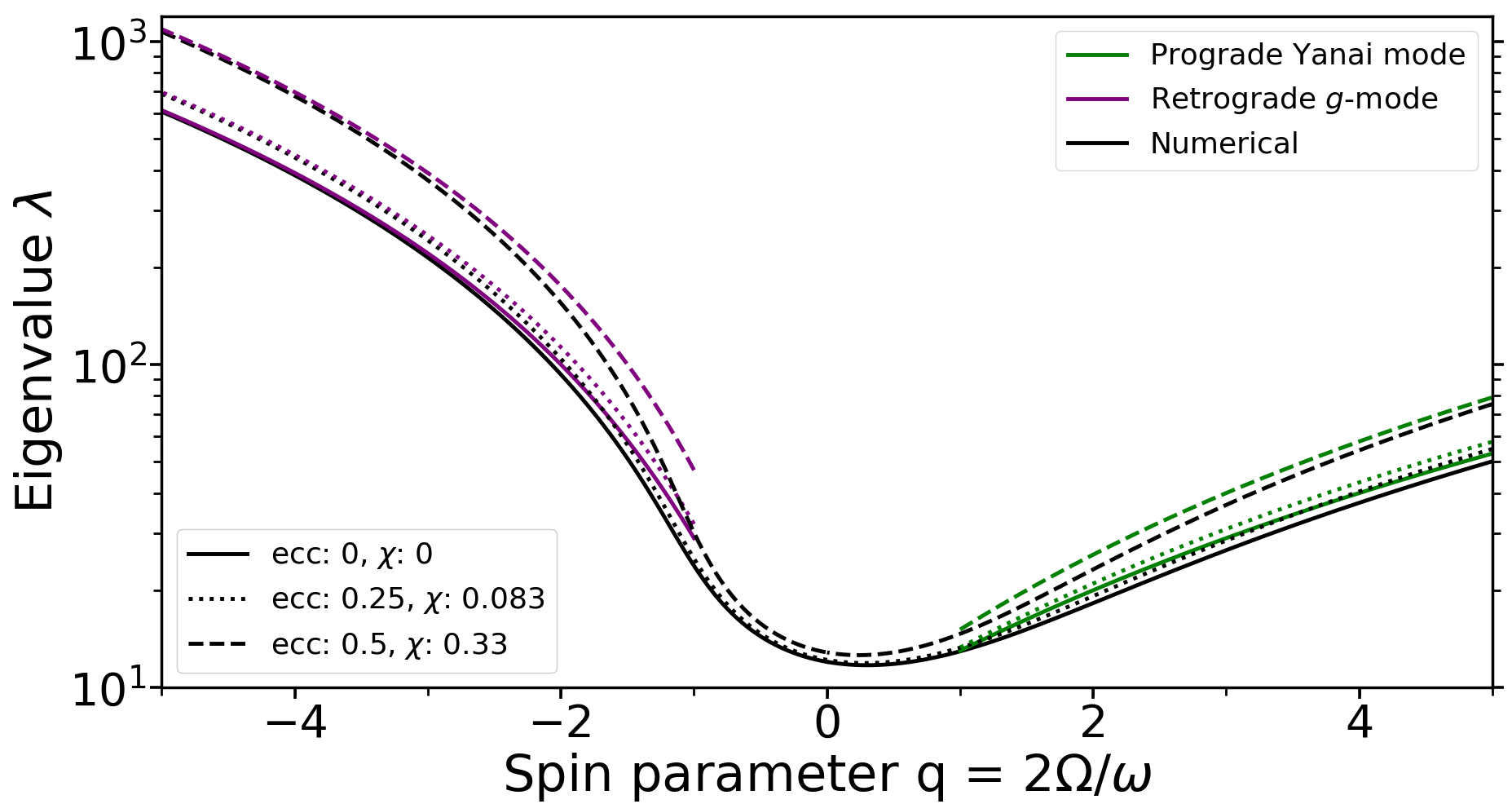}
    \caption{A comparison between the numerically calculated (black) and approximate (in colour) solutions to the $m=-2$ prograde and retrograde waves with $k=1$ for different eccentricities. Purple corresponds to a retrograde $g$-mode and green to a prograde Yanai mode. The solid lines correspond to $e=0$, the dotted lines to $e=0.25$ and the dashed lines to $e=0.5$, with the relation between $e$ and $\chi$ as given by Equation (\ref{eq:realistic_chi_e_rel}). The analytic approximations are given by Equation (\ref{eq:lambda_gmodes}) for both modes.}
    \label{fig:k1proretro}
\end{figure}

\begin{figure}
	\includegraphics[width=\linewidth]{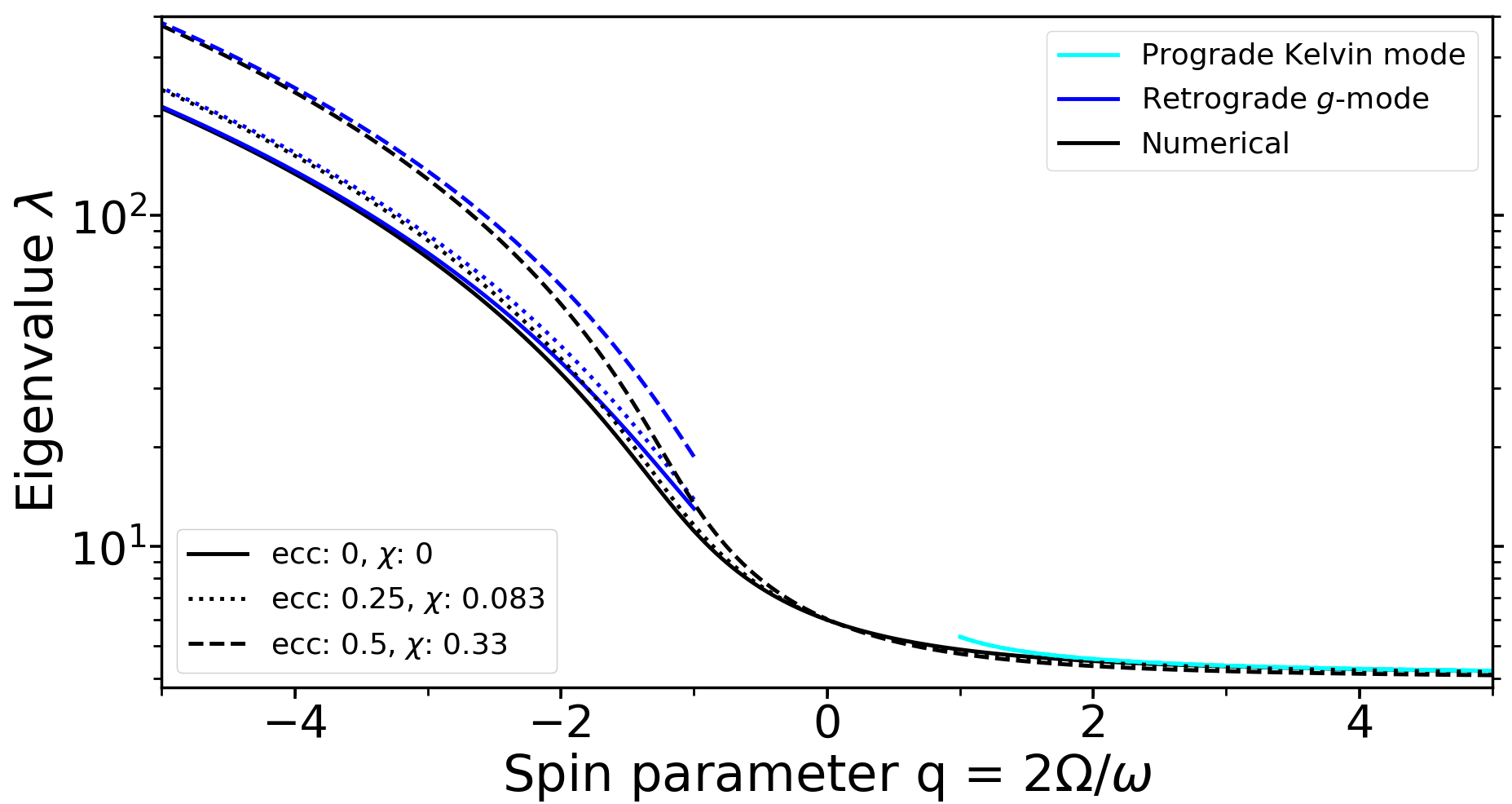}
    \caption{A comparison between the numericallu calculated (black) and approximate (in colour) solutions to the $m=-2$ prograde and retrograde waves with $k=0$ for different eccentricities. Dark blue corresponds to a retrograde $g$-mode and light blue to the prograde Kelvin mode. The solid lines correspond to $e=0$, the dotted lines to $e=0.25$ and the dashed lines to $e=0.5$, with the relation between $e$ and $\chi$ as given by Equation (\ref{eq:realistic_chi_e_rel}). The analytic approximations are given by Equations (\ref{eq:lambda_gmodes}) for the retrograde (negative $q$) and (\ref{eq:lambda_kelvinmodes}) for the prograde mode, respectively.}
    \label{fig:k0proretro}
\end{figure}

\begin{figure}
	\includegraphics[width=\linewidth]{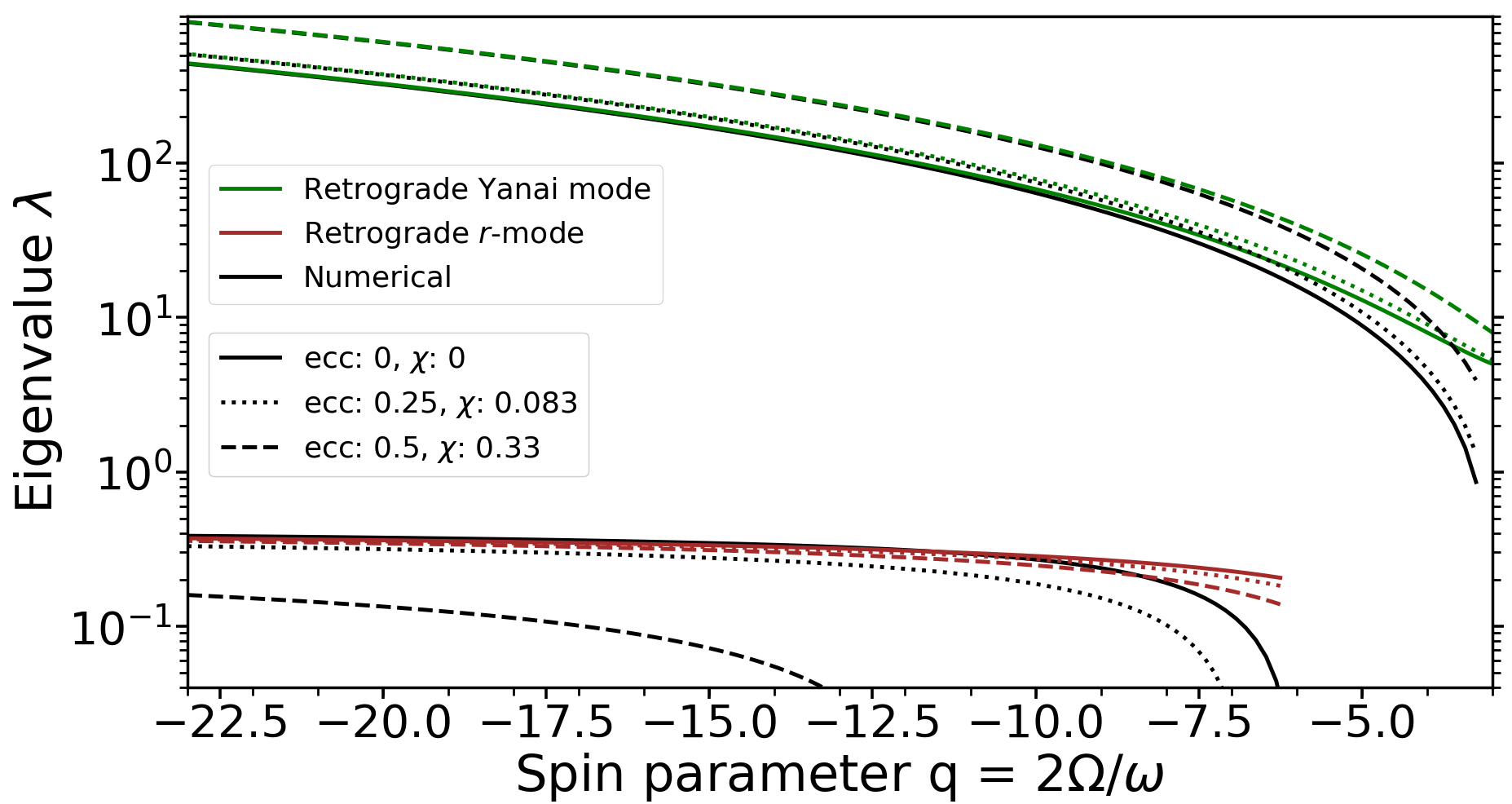}
    \caption{A comparison between the numerically calculated (black) and approximate (in colour) solutions to the $m=-2$ retrograde Yanai ($k=-1$) and $r$-mode ($k=-2$) waves for different eccentricities of $e=0$ (solid lines), $e=0.25$ (dotted) and $e=0.5$ (dashed). The Yanai mode (green) is approximated by Equation (\ref{eq:lambda_gmodes}) and matches well for $|q|>10$. For increasing eccentricity, the analytic approximations of the $r$-mode (brown) do not match with the numerical solution. The approximation for the eigenvalues are calculated using Equation (\ref{eq:lambda_gmodes}) for the Yanai mode and Equation (\ref{eq:lambda_rmodes}) for the $r$-mode. The relation between $e$ and $\chi$ is given by Equation (\ref{eq:realistic_chi_e_rel}).}
    \label{fig:knegretro}
\end{figure}

\subsection{Comparison to numerical solutions}

With new asymptotic approximations to eigenvalues including oblateness, we now investigate how well they compare to numerical solutions, and how increasing oblateness changes the values compared to their spherical counterparts as calculated by \citet{townsend2003asymptotic}. We use a shooting method to calculate the eigenvalues and eigenfunctions, and use the same normalisation condition as in \citet{townsend2003asymptotic}.

In the non-rotating limit $P_p$ reduces to the associated Legendre polynomials $P^m_l$ \citep{lee1997low} (in the case of $g$-modes, as $r$-modes do not exist in non-rotating systems). For these polynomials, $m$ is the azimuthal order and the integer $l\geqslant0$ is the harmonic degree. However, with this classification it is impossible to describe retrograde Yanai modes and $r$-modes, as those do not exist in the non-rotating limit. \citet{lee1997low} came up with an alternative classification scheme which uses a unique integer index $k$ for each solution of the tidal equations. Positive and zero values of this index indicates that the mode possesses a counterpart of harmonic degree $l=|m|+k$, in the limit of no rotation, while negative values of $k$ denote $r$-modes and retrograde Yanai modes.  Furthermore, odd values of $k$ denote odd-parity modes while even values of $k$ correspond to even-parity modes.

It is straightforward to relate the $k$-index used by \citet{lee1997low} with the $s$-index which appears in the approximations of Equations (\ref{eq:lambda_gmodes}) and (\ref{eq:lambda_rmodes}), which was also used by \cite{townsend2003asymptotic}. For prograde modes, the relation is given by $s=k-1$, while for the retrograde modes there is a split, as the relation is $s=-k-1$ for the $r$-modes and $s=k+1$ for the $g$-modes; both of these result in the correct relation for the retrograde Yanai mode (which has $k=-1\Leftrightarrow s=0$). The advantage of using the $k$-index over $s$ is that the $k$-indexing scheme can uniquely identify all modes, while for the $s$ scheme additional information would be necessary in order to differentiate between $g$-modes and $r$-modes.

In Figures \ref{fig:k2proretro}, \ref{fig:k1proretro}, \ref{fig:k0proretro} and \ref{fig:knegretro} we show the numerical solution and analytic approximation to the eigenvalues for the $m=-2$ and $k=2$, $k=1$, $k=0$ and $k=-1,-2$ modes. In each of the Figures, three different eccentricities $e=0,0.25,0.5$ are used. The other parameter we have to choose is $\chi$, for which the relation in Equation (\ref{eq:realistic_chi_e_rel}) is used. From Equation (\ref{eq:lambda_gmodes}) however, it can be seen that if $|q|$ is large then the impact of $\chi$ will be marginal compared to the impact of $e$.

For the $k=2$ modes (which are $g$-modes) in Figure \ref{fig:k2proretro} it can be seen that beyond $|q|=5$ the analytic approximation matches well with the numerical solution for all eccentricities. For both the prograde and retrograde $g$-modes, the eigenvalues increase with $q$. Higher eccentricities (dotted and dashed lines) results in a steeper slope for $\lambda$ against $q$ ($\lambda$ increases more rapidly when $e$ goes from $0.25$ to $0.5$ than it does from $0$ to $0.25$).
For the retrograde $g$-mode in Figure \ref{fig:k1proretro} the same effects can be observed. There is a good match between the numerical and approximate values at $|q|=5$. Eigenvalues scale with $e$ as predicted by the asymptotic approximation. On the prograde side of the Figure however, it can be seen that the Yanai mode converges more slowly to the approximation of Equation (\ref{eq:lambda_gmodes}), but it does also follow the same trend of scaling with $e$.

In Figure \ref{fig:k0proretro}, the retrograde $g$-mode repeats the behaviour of the other $g$-modes seen for $k=2,1$. At $|q|=5$ the analytic approximations and the numerical solution agree, and for increasing $e$ the eigenvalues increase.
For the retrograde Yanai and the $r$-modes, we need to look at higher values of $|q|$ as there are no positive solutions for lower values of $q$ for these modes. In Figure \ref{fig:knegretro} a comparison is shown between the numerical solutions and approximations to the eigenvalues. For the retrograde Yanai mode, the numerical solutions and analytic approximations agree for values of $|q|>10$, but converge quite slowly.

Unlike the other wave families, the analytic approximation for the $r$-mode does not match with the numerical solution for higher eccentricity. It can be seen that for the zero eccentricity (solid lines) the approximation and numerical solutions do converge to the same value for high $|q|$, and the analytic approximation changes little for increasing $e$. This is not the same behaviour shown by the numerical solutions. A higher value of $|q|$ is required to find positive eigenvalues, and the asymptotic limit which these eigenvalues approach is much smaller for increasing $e$.

We test the effect of $\chi$ on the eigenvalues by using a different relation between $e$ and $\chi$, such as $\chi=2e^2$. For the $g$-modes and Yanai modes, any differences are negligible, which is not unexpected given that terms with $\chi$ scale with $q$, while terms with $e$ scale with $q^2$ in the approximation for $\lambda$ in Equation (\ref{eq:lambda_gmodes}). We show the impact for the $r$-mode in  Section \ref{sec:NS}.

\subsection{Kelvin modes}

The final family of modes which we will investigate are the Kelvin modes. In the spherical case, the eigenvalue of these modes may be approximated by $\lambda \approx m^2$ in the limit of $q\rightarrow\infty$ (arbitrary but rapid rotation). We make the same approximations associated with equatorial trapping for these modes as was performed for the $g$-modes and $r$-modes. However, we cannot drop the $\mu^2$ dependence on the right hand side of Equation (\ref{eq:lammsq_Ptheta_approx}) since $\lambda$ and $m^2$ are not appreciably different. Furthermore, this fact implies that $P_p$ must be much larger than $P_\theta$ since we require the term on the right hand side of the equation ($\sim \lambda \mu^2 P_p$) to be the same order as $m q \mu P_\theta$. This leads to the set of equations:
\begin{equation}
	\bigg( \dfrac{\text{d}}{\text{d}\mu} - mq\mu \bigg) P_p = 0,
    \label{eq:lammsq_Pp_approx}
\end{equation}
\begin{equation}
	\bigg( \dfrac{\text{d}}{\text{d}\mu} + mq\mu \bigg) P_\theta - P_\theta\dfrac{\text{d}\ln{g}}{\text{d}\mu} = \sigma \bigg[ \lambda(1-\mu^2) - m^2 \bigg]P_p,
    \label{eq:lammsq_Ptheta_approx}
\end{equation}
Equation (\ref{eq:lammsq_Pp_approx}) is unchanged from spherical geometry, allowing the same solution:
\begin{equation}
    P_p(\mu) = e^{mq \mu^2 / 2}.
    \label{eq:Pp_curvi_tauform}
\end{equation}
We require $mq<0$ (prograde motion) since the mode must decay to zero at the pole. In the astrophysical literature they have, because of this, been referred to as low-frequency prograde waves \citep{unno1989nonradial}. Applying the solution for $P_p$ to Equation (\ref{eq:lammsq_Ptheta_approx}) we find the following first order differential equation:
\begin{equation}
	\dfrac{\text{d}}{\text{d}\mu}\bigg( e^{mq \mu^2 / 2}P_\theta \bigg) - e^{mq \mu^2 / 2}P_\theta \dfrac{\text{d}\ln{g}}{\text{d}\mu} = \sigma \bigg[\lambda( 1 - \mu^2 ) - m^2 \bigg]e^{mq \mu^2}.
    \label{eq:Ptheta_kelvinapprox_withgrav}
\end{equation}
This equation can be solved using integrating factors. However, we find that the solution for the eigenvalues in spherical geometry is valid for even large values of eccentricity. This solution is:
\begin{equation}
	\lambda = m^2\dfrac{2mq}{2mq+1},
    \label{eq:lambda_kelvinmodes}
\end{equation}
which satisfies the initial assumption that $\lambda \approx m^2$.

In Figure \ref{fig:k0proretro} the numerical solutions are shown for different eccentricities $e$ and compared to the analytic approximation of Equation (\ref{eq:lambda_kelvinmodes}). For all eccentricities the analytic approximation is a good match, which shows that the Kelvin modes on  oblate spheroids can still be approximated using the spherical analysis. For increasing eccentricities a slight decrease in the eigenvalue can be observed for this mode, although the difference from the spherical eigenvalue is small. 

We test the effect of $\chi$ on the eigenvalues by using a different relation between $e$ and $\chi$, namely $\chi=2e^2$. For the Kelvin mode, the numerical solutions decrease slightly more for this new relation but still match well to the spherical approximation.

\begin{figure}
	\includegraphics[width=\linewidth]{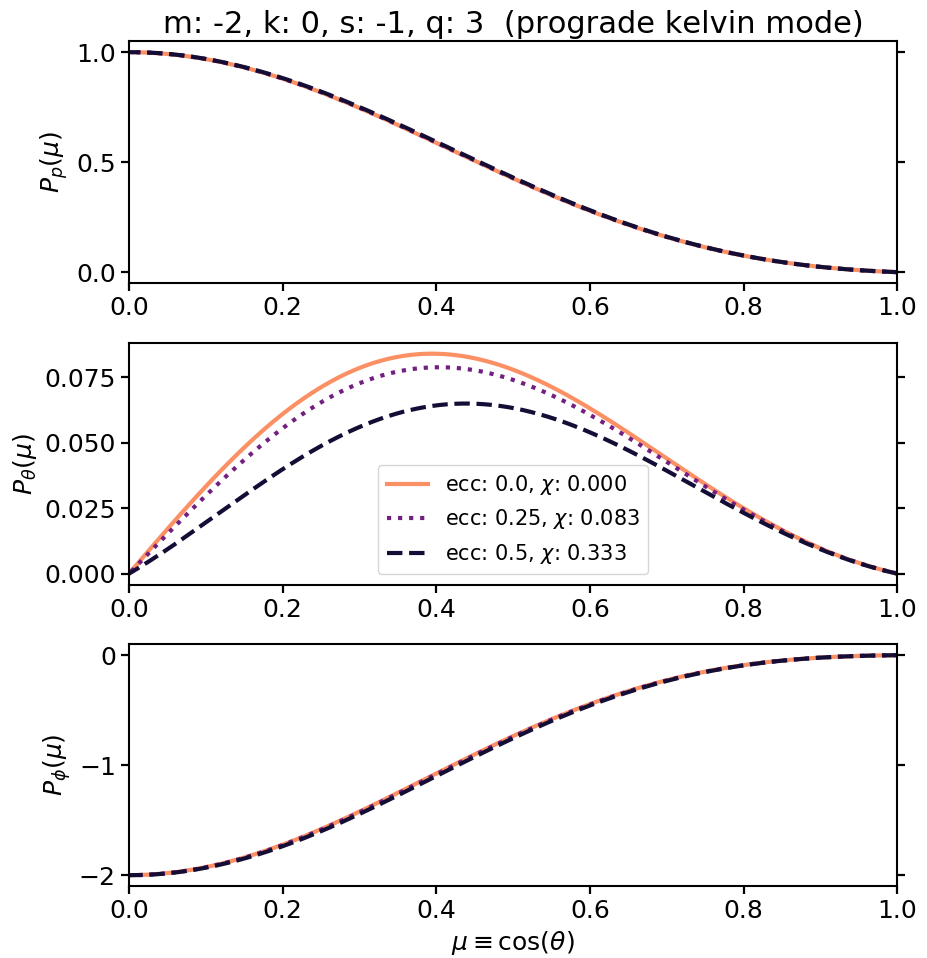}
    \caption{A comparison between the numerically calculated eigenfunctions $P_p$, $P_\theta$ and $P_\phi$ for the $m=-2$ Kelvin mode at $q=3$, for different eccentricities of $e=0$ (solid lines), $e=0.25$ (dotted) and $e=0.5$ (dashed). For more eccentric systems, the equatorial confinement of the Kelvin mode is weaker. The peak of $P_\theta$ is found at higher latitudes. This behaviour is only observed for the Kelvin mode, as the other modes all have stronger equatorial confinement for more eccentric systems. The relation between $e$ and $\chi$ is given by Equation (\ref{eq:realistic_chi_e_rel}).}
    \label{fig:kelvin_hough}
\end{figure}

The eigenfunctions of the Kelvin wave depend on the oblateness in an interesting manner. Figure \ref{fig:kelvin_hough} shows that for increasing eccentricity the peak of the $P_\theta$ function moves to higher latitudes, indicating that the wave is less confined to the equatorial region for more oblate systems. This sets the Kelvin modes apart from the $g$-modes and Yanai modes, as those become more equatorially confined in more eccentric systems. We find that even for the most oblate systems, the new term in equation (\ref{eq:lammsq_Ptheta_approx}), related to gravitational variation across the surface of the star, is small relative to the other terms which appear in that equation. The lack of influence of gravitational variation on the Kelvin wave is also reflected in the minor differences in eigenvalues when including this effect.

\section{Waves on Neutron Star Oceans} \label{sec:NS}

\begin{figure}
  \includegraphics[width=\linewidth]{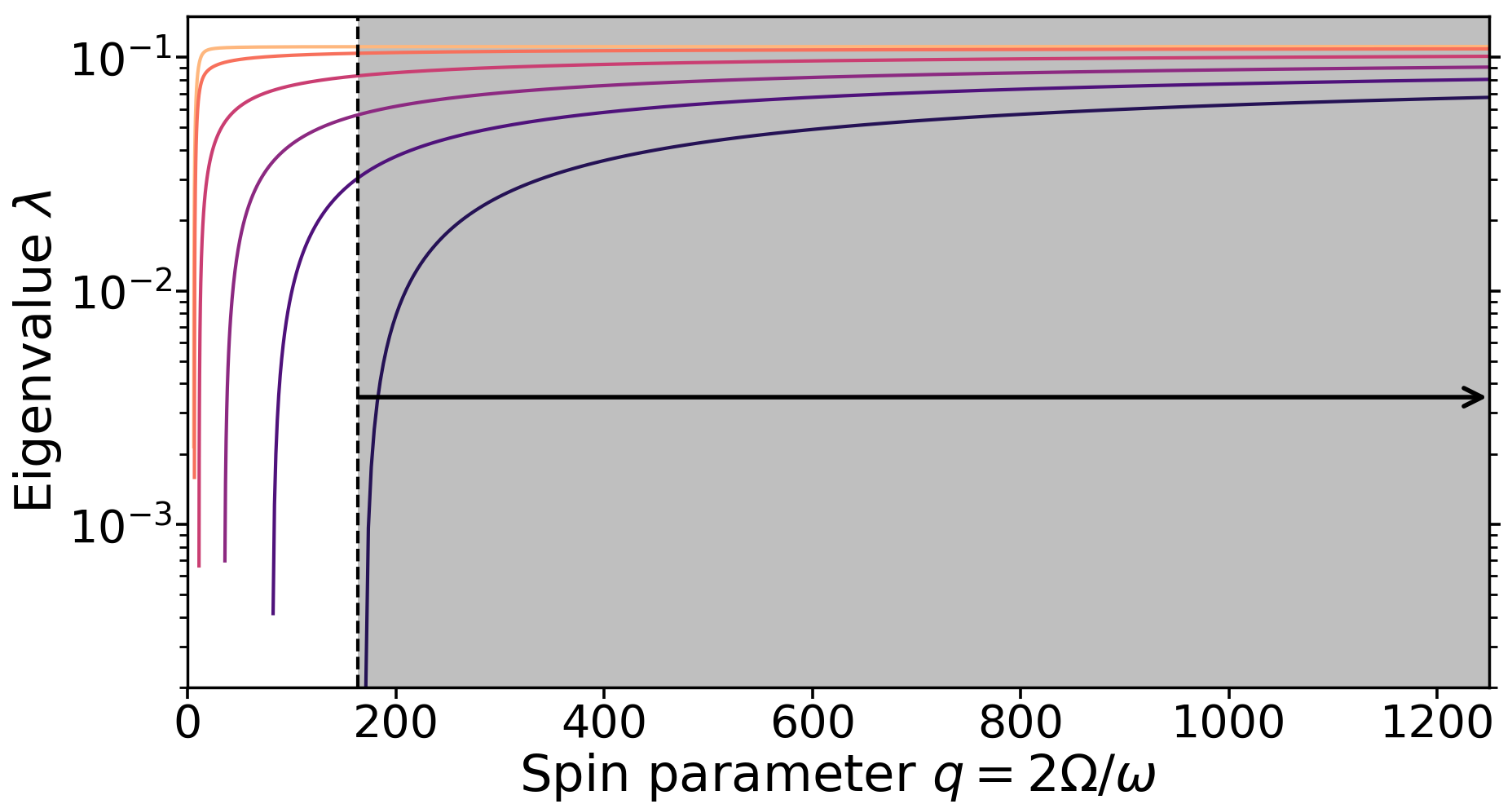}
      \caption{Numerically calculated eigenvalues for several values of eccentricity ($e=0, 0.125, 0.25, 0.35, 0.425, 0.5$ from top to bottom going light to dark) for the $m=1$, $k=-2$ $r$-mode. Equation (\ref{eq:realistic_chi_e_rel}) is used to relate $e$ and $\chi$. The grey region reflects the values of $q$ which are of interest for the study of thermonuclear burst oscillations as described in Section \ref{sec:NS}.}
    \label{fig:rmode_eigenvalue_slowrel}
\end{figure}

\begin{figure}
	\includegraphics[width=\columnwidth]{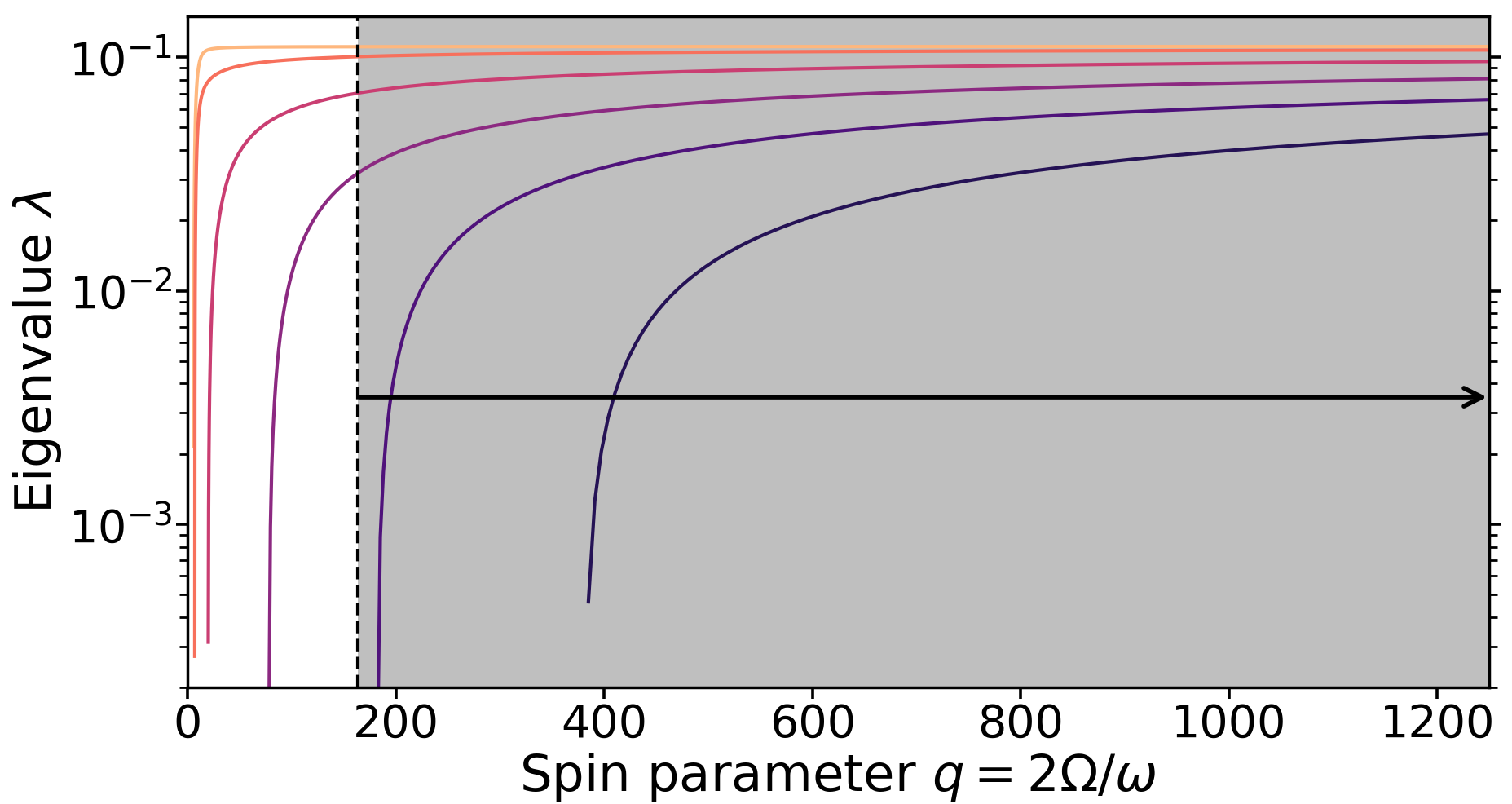}
    \caption{Same $m=1, k=-2$ $r$-mode as Figure \ref{fig:rmode_eigenvalue_slowrel}, using $\chi = 2 e^2$ instead of Equation (\ref{eq:realistic_chi_e_rel}).}
    \label{fig:rmode_eigenvalue_highrel}
\end{figure}

\begin{figure}
\includegraphics[width=\columnwidth]{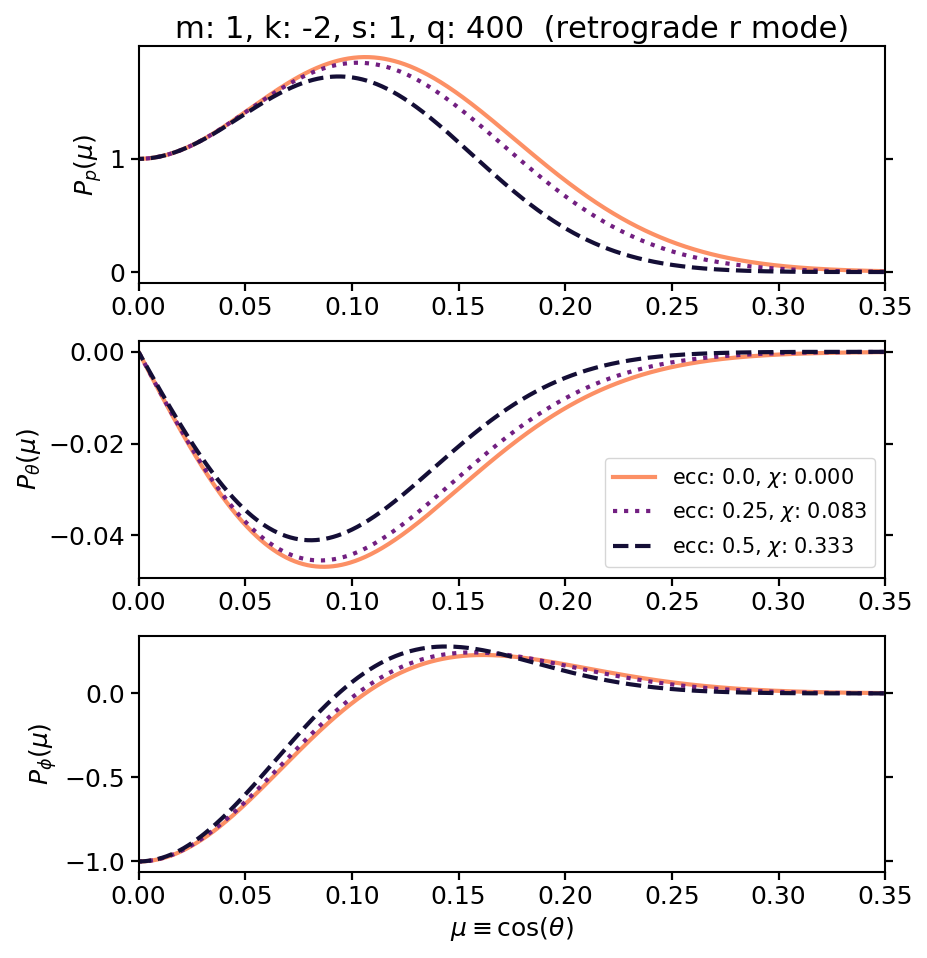}
\caption{
The eigenfunction of the $m=1$, $k=-2$ retrograde $r$-mode with $q=400$ over the range $\mu = 0$ (the equator) to $\mu=0.35$ ($\approx 20\degr$ in latitude) for three different eccentricities. The relation between $e$ and $\chi$ is given by Equation (\ref{eq:realistic_chi_e_rel}). For larger eccentricities, the waves are more confined to the equator and the peak of the wave functions also moves towards lower latitudes.}
\label{fig:hough_funcs_rmode}
\end{figure}

Rotationally induced oblateness is relevant for neutron stars, in particular for surface modes which may be responsible for the phenomenon of thermonuclear burst oscillations. The most likely mode candidate to explain the observational properties of thermonuclear burst oscillations, suggested by \citet{heyl2004r}, is the $m=1$ $r$-mode with a small number of latitudinal nodes ($k = -2, -3$). The rotating frame frequency of one of these modes must be small, $\omega / 2 \pi \sim 1 - 3\,$Hz \citep[see][for further details]{heyl2004r}, while the angular velocity of the source is assumed to be large, between $245 - 620\,$Hz \citep{watts2012thermonuclear}. These two constraints mean that the spin parameter $q$ falls in the range $161 - 1240$.

Sections \ref{sec:gravintro} and \ref{ssec:eigenvalues} suggested a model for the variation in gravitational acceleration with latitude, tailored to neutron stars based on calculations for a range of stellar masses, radii, equations of state and rotation rates \citep{algendy2014universality}. We continue to use this model. Section \ref{sec:approximations} showed that the asymptotic approximations deviate significantly from the numerical solutions for the $r$-modes, and as such, for the rest of this analysis we rely purely on numerical solutions.

In Figure \ref{fig:rmode_eigenvalue_slowrel} the eigenvalues are shown for a range of values of $e$, the region where we expect thermonuclear burst oscillation sources to be present is marked in grey. We wish to investigate the effect of a larger change in gravitational acceleration across the surface of the neutron star. To do this, we use the relation $\chi=2e^2$ instead of $\chi=4/3e^2$ and show results in Figure \ref{fig:rmode_eigenvalue_highrel}. This different relation means that for the same value of $e$, we now have a higher value of $\chi$, so the variation of the gravitation acceleration between the equator and the poles in greater than was shown in Figure \ref{fig:rmode_eigenvalue_slowrel}. 

A larger eccentricity increases the value of $q$ for which the eigenvalues become positive. Using larger values of $\chi$ this effect becomes more extreme. Figure \ref{fig:rmode_eigenvalue_slowrel} shows that for the relation $\chi = 4/3 e^2 $, this transition occurs at $q = 11$ for $e = 0.25$, and $q = 170$ for $e=0.5$. In Figure \ref{fig:rmode_eigenvalue_highrel}, with the relation $\chi = 2 e^2$, it occurs at $q = 20$ for $e = 0.25$, and $q = 383$ for $e=0.5$. In the spherical case, this transition occurs at $q = 6$ \citep{lee1997low}.

The analytic approximation for the eigenvalue of $r$-modes, Equation (\ref{eq:lambda_rmodes}), indicates that the $r$-modes in eccentric systems should reach the same asymptotic value as those in spherical systems. The analytic approximations also predict that the eigenvalue of an $r$-mode in an oblate system is always smaller than that of a corresponding spherical system for the same value of $q$. This prediction, however, is much smaller than shown in the numerical results. The rate at which the eigenvalue converges to an asymptotic limit is decreased significantly when including oblateness. For the most eccentric systems, the modes have not yet approached their asymptotic limit by $q=1250$, while in spherical geometry the asymptotic limit is reached at $q \approx 20$, a much smaller value. This discrepancy either indicates that the asymptotic limit differs between spherical and oblate systems, or that oblateness changes the rate at which the eigenvalue converges to an asymptotic limit more than is predicted by the analytic approximations. Considering the very great discrepancy between the eigenvalues calculated using different relations between $\chi$ and $e$, it is possible that the approximate form for gravitational acceleration across the surface of the star (the one used to make (\ref{eq:Ptheta_ODE_gravapproximated})) was not a reasonable choice, leading to a less accurate formula for the eigenvalues of $r$-modes.

It can be seen that for the most eccentric system ($e = 0.5$) the range of thermonuclear burst oscillation sources is close to the point at which $\lambda$ becomes positive. This eccentricity, however, will likely only be reached by the most rapidly rotating neutron stars at values of $|q|\gg 245$. Since $\omega$ (and therefore $|q|$) changes throughout the course of a burst, $\lambda$ will also change during a burst, which was not the case when using the spherical eigenvalue.

Obtaining the correct eigenvalues is important for calculating the frequency of the wave when including the radial structure of the mode. Equation (3) of \citet{piro2005surface} approximates the dependence of $\omega$ on the eigenvalue as $\lambda^{1/2}$. By self-consistently solving $\omega=2\Omega/q = C\lambda(q)^{1/2}$, where $C$ is a constant, we can investigate the change in frequency of the wave when including oblateness. We consider the eigenvalues at $q=250$ and $1250$ for $e=0.25, 0.5$ as shown in Figure \ref{fig:rmode_eigenvalue_slowrel}. We find that for $e=0.25$ the wave frequency would be reduced by $5\%-11\%$, while for $e=0.5$ the wave frequency would be reduced by $25\%-60\%$.

Figure \ref{fig:hough_funcs_rmode} shows eigenfunctions of the $r$-mode for several values of eccentricity calculated using $q=400$ and Equation (\ref{eq:realistic_chi_e_rel}) to relate $e$ and $\chi$. For higher eccentricities, each component of the eigenfunction is more confined to the equator. The area over which the $P_p$ component of the eigenfunction is appreciably different from zero ($P_p > 0.01 P_{p, \mathrm{max}}$) changes by $\approx 16\%$ when the eccentricity is changed from $0$ to $0.5$. Using the relation $\chi=2e^2$, this area shrinks by $20\%$. Greater equatorial confinement should result in a reduced pulsed amplitude and therefore the observed thermonuclear burst oscillation amplitudes since the contrasting pattern of the wave is contained in a smaller area \cite{heyl2005r}.

\section{Discussion} \label{sec:discussion}

In this work we investigated the impact of rotationally induced oblateness on waves that exist in a thin fluid layer on the surface of a spheroid. Compared to their counterparts on the surface of a slowly rotating sphere we found the eigenfunctions and the eigenvalues change. Each family of modes is modified in a different manner. These results were applied to the case of accreting neutron stars and specifically the phenomenon of thermonuclear burst oscillations, where rotationally induced oblateness is expected to be relevant given the high spin rates. We investigate the $m=1$ $r$-mode and find that for more oblate systems, equatorial confinement increases and the wave frequency decreases.

The numerical solutions and analytic approximations for $m = -2$ modes agree well with each other for the $g$-modes at $|q|>4$ and for the Yanai modes at $|q|>10$. No new approximations are found for the Kelvin modes since the approximations from spherical geometry match well the solutions calculated including oblateness. For the $r$-modes however, the analytic approximations no longer match the numerical eigenvalues, especially for large oblateness and higher variations in gravitational acceleration across the surface of the spheroid.

We find that the eigenvalues of the $g$-modes, Yanai modes and $r$-modes are altered by the presence of oblateness and do not depend particularly strongly on the model for gravitational acceleration across the surface of the star; however, they do depend on the degree of oblateness. 
This effect is predicted by the asymptotic approximation Equation (\ref{eq:lambda_gmodes}), where the leading term scales with $e$, the eccentricity, and not with $\chi$, the parameter that characterises the change in gravitational acceleration across the surface of the star. 
The $r$-mode, on the other hand, depends strongly on $\chi$ as can be seen from the differences in eigenvalues in Figures \ref{fig:rmode_eigenvalue_slowrel} and \ref{fig:rmode_eigenvalue_highrel}.
For the same values of $e$ but higher $\chi$, the eigenvalues become smaller and only become positive towards higher values of the spin parameter $|q|$. This behaviour is not predicted from the asymptotic approximation, but Figure \ref{fig:knegretro} shows that the approximation no longer works when more eccentric systems are considered. 
For the Kelvin mode, the eigenvalues decrease slightly for larger oblateness but the effect is small enough that the spherical approximations are still valid.
However, these modes do have one property that distinguishes them from the other families of modes; they become less equatorially confined for larger oblateness which is precisely the opposite of what is seen for the $g$-mode, Yanai mode and $r$-mode (see Figures \ref{fig:kelvin_hough} and \ref{fig:hough_funcs_rmode}).

For the $m=1$ $r$-mode we investigated how the eigenvalues and eigenfunctions change due to oblateness. This has implications for the wave model for thermonuclear burst oscillations. We find that for systems with an eccentricity between $0.25 - 0.5$, the frequency of the wave drops by $5-60\%$, while the area in which wave amplitude is significantly different from zero shrinks by $\sim20\%$. A less equatorially confined pattern on the surface of the neutron star should lead to a larger burst oscillation amplitude; our findings suggest that for rapidly rotating (and thus more oblate) neutron stars, thermonuclear burst oscillations might (if caused by this mechanism) be more difficult to detect, as they would exhibit lower pulsed amplitudes. This might help explain why we do not see burst oscillation sources which spin faster than $620\,$Hz (if this is not due to the lack of more rapidly spinning neutron stars, \citealt{manchester2005australia}). Note however that thermonuclear burst oscillation amplitude also depends on other factors such as the accretion rate \citep{Ootes2017accretion}, something that might confound efforts to isolate the effects of rotation rate on amplitude in the current burst oscillation data set \citep{bilous2019uniform}.

It is important to note that our results for the $r$-modes depend strongly on the parametrisation of the gravitational acceleration across the neutron star surface. With our simple parametrisation, Equation (\ref{eq:grav_eq}), significant differences are present when increasing the difference in the gravitational acceleration at the pole and the equator, changing the asymptotic limit of the eigenvalues by up to a factor of 2 and the area over which the eigenfunctions significantly differs from zero by as much as $20\%$. Our parametrisation was tailored for neutron stars specifically and based on the work of \citet{algendy2014universality}, but other parametrisations might fit better for applications other than for neutron stars.

Inferring the rotation rate of neutron stars from the phenomenon of thermonuclear burst oscillations has implications for continuous gravitational wave searches \citep{watts2008detecting} and modelling the spacetime surrounding a neutron star \citep{riley2018parametrized}. Any viable model for this phenomenon needs to provide a robust relationship between the burst oscillation frequency and the rotation rate of the star; in the mode model, this is the mode frequency in the rotating frame. Our study shows that the mode frequency could decrease by as much as $\sim 60\%$ due to oblateness for the most rapidly rotating systems. \citet{maniopoulou2004traditional} estimated that general relativistic effects can decrease the frequency of modes on the surface of a neutron star by $\sim20\%$, and \citet{chambers2019relativistic} found similar reductions in frequency for $r$-modes when including the radial structure. Other effects are also known to be important, such as nuclear burning throughout the course of the burst \citep{chambers2019burning}, and perhaps magnetic fields \citep{heng2009magnetohydrodynamic}. A complete mode model for thermonuclear burst oscillations should eventually include oblateness alongside all of these other effects.

\section*{Acknowledgements}

The authors acknowledge support from ERC Starting Grant No. 639217 CSINEUTRONSTAR (PI: Watts). This work benefited from discussions at the `Bursting the Bubble' Lorentz Center workshop. The authors would like to express their gratitude to the referee Richard Townsend for his helpful and insightful comments, in particular with regards to the impacts of the wave frequency, and in general to improve the clarity of this work.




\bibliographystyle{mnras}
\bibliography{burning_oceans} 









\bsp	
\label{lastpage}
\end{document}